\documentclass[a4paper]{article}
\input{epsf.sty}
\begin{document}

\begin{center}
{\LARGE A relation among the effective nucleon mass, the incompressibility and the effective $\sigma$-meson mass in nuclear matter}
\end{center}

\centerline{Yoshitaka Iwasaki, Hiroaki Kouno, Akira Hasegawa and Masahiro Nakano$^*$}

\centerline{Department of Physics, Saga University Saga 840, Japan}

\centerline{*University of Occupational and Environmental Health, Kitakyushu 807, Japan}

\centerline{Abstract}
A relation among the effective nucleon mass $M^*$, the incompressibility $K$ and the effective $\sigma$-meson mass $m_{\rm s}^*$ in nuclear matter is studied by using the relativistic nuclear model. 
We found that there is a strong correlation between $M^*$ and $m_{\rm s}^*$, while there is only a weak correlation between $K$ and $m_{\rm s}^*$. 
At the normal density, $m_{\rm s}^*$ is smaller than the one at zero density, if $M^*$ is smaller than 0.8 times of the nucleon mass at zero density. 
It is also found that the off-shell effective mass $\mu_{\rm s}^*$ is related directly to $K$ and $M^*$ at the normal density. 

\bigskip
\section{Introduction}
Recently, the $\pi$-$\pi$ scattering phase shift is reanalyzed and the existence of the light iso-singlet scalar $\sigma$-meson is strongly suggested. \cite{rf:A1} 
The similar results are also obtained by reanalyzing the $\pi^0\pi^0$ mass spectra and angular distributions around $K\bar{K}$-threshold and at 1.5GeV in ${p\bar{p}}$(at rest) $\rightarrow$3$\pi^0$. \cite{rf:Ishida} 

Although the existence of the $\sigma$-meson is not still established, this meson play an important role for the nuclear matter properties in the quantum hadrodynamics(QHD). 
For example, the nuclear saturation properties are realized by a balance of attractive effects of the $\sigma$-meson and repulsive effects of the $\omega$-meson. \cite{rf:Walecka}

The effective self-interactions (or potentials) of $\sigma$-meson play an important role in determining the effective nucleon mass and the incompressibility of the nuclear matter. 
\cite{rf:Boguta} 
Inversely, in QHD, the properties of the effective potentials in the symmetric nuclear matter are almost determined if the values of the effective nucleon mass $M_0^*$ at the normal density and the incompressibility $K$ are given as input parameters. 
The effective $\sigma$-meson mass is also determined if the values of these two quantities are given, since it can be defined as a second derivative of the effective potential with respect to the $\sigma$-meson field. 
In this paper, we study the relation among the effective nucleon mass, the incompressibility and the effective $\sigma$-meson mass within the framework of QHD. 
\section{Formalism}
We use the relativistic Hartree approximation (RHA) \cite{rf:Chin} based on the $\sigma$-$\omega$ model. \cite{rf:Walecka} 
The Lagrangian density is composed of three fields, the nucleon $\psi$, the scalar $\sigma$-meson $\phi$ and the vector $\omega$-meson $V^\mu$, and is given by 
\begin{eqnarray}
 L &=& \bar \psi (i\gamma _\mu  \partial ^\mu   - M + g_{\rm s} \phi  - 
g_v \gamma _\mu  V^\mu  )\psi  
+\frac{1}{2}\partial _\mu  \phi \partial ^\mu  \phi  - 
\frac{1}{2}\mu_{\rm s}^2 \phi ^2 
 {\rm       }\nonumber \\ &-& \frac{1}{4}F_{\mu \nu } F^{\mu \nu }  + 
\frac{1}{2}\mu_{\rm v}^2 V_\mu  V^\mu  {\rm  }+\sum_{n=0}^4C_n\phi ^4; ~~~~~~F_{\mu\nu}=\partial_\mu V_\nu-\partial_\nu V_\mu
\label{eq:1} 
\end{eqnarray}
where $M$, $\mu_{\rm s}$, $\mu_{\rm v}$, $g_{\rm s}$, $g_{\rm v}$ and $C_n$ are constant parameters. 
The last term in (\ref{eq:1}) is a counter term and $C_n$ is determined by the phenomenological renormalization conditions at zero baryon density, namely, 
\begin{equation}
C_n={1\over{n!}}{\partial^n\over{\partial <\phi >^n }}U^{\rm V}_{\rm 1-loop}(<\phi> )~~~~~~(n=0,1,2,3,4), 
\label{eq:ad3}
\end{equation}
where $U^{\rm V}_{\rm 1-loop}$ is the unrenormalized 1-loop effective potential induced by the vacuum fluctuations and $<\phi >$ is the ground state expectation value of the $\sigma$-meson field. 

Replacing the meson fields by their ground state expectation values $<\phi >$ and $<V^0>$, we obtain the equation of motions for $< \phi >$ and for $< V^0 >$, namely, 
\begin{equation}
\rho_{\rm s}={1\over{C_{\rm s}^2}}g_{\rm s}<\phi> ={1\over{C_{\rm s}^2}}(M-M^*) 
\label{eq:ad1}
\end{equation}
and 
\begin{equation}
\rho ={1\over{C_{\rm v}^2}}g_{\rm v}<V^0> , 
\label{eq:ad2}
\end{equation}
where $M^*(\equiv M-g_{\rm s}<\phi >)$, $\rho_{\rm s}$ and $\rho$ are the effective nucleon mass, the scalar density and the baryon density, respectively, and $C_{\rm s}=g_{\rm s}/\mu_{\rm s}$ and $C_{\rm v}=g_{\rm v}/\mu_{\rm v}$. 
The scalar density $\rho_{\rm s}=<\bar{\psi}\psi >$ and the baryon density $\rho =<\bar{\psi}\gamma_0\psi >$ are given by 
\begin{eqnarray}
\rho_{\rm s}(k_F,<\phi >)&=&{\lambda\over{2\pi^2}}M^*\left[ k_FE_F^*-M^{*2}\ln{\left( {k_F+E_F^*\over{M^*}}\right)}\right] 
\nonumber\\
&-&{1\over{g_{\rm s}}}{dU^{\rm V,R}_{\rm 1-loop}\over{d<\phi >}}\equiv \rho_{\rm s}^{\rm D}(k_F,<\phi >)+\rho_{\rm s}^{\rm V}(<\phi >)  
\label{eq:ad6}
\end{eqnarray}
and   
\begin{equation}
\rho (k_F) ={\lambda\over{3\pi^2}}k_F^3, 
\label{eq:ad7}
\end{equation}
where $\lambda$ is the number of degree of freedom of nucleons, $k_F$ is the Fermi momentum, $E_F^*=\sqrt{k_F^2+M^{*2}}$, respectively. 
The $\rho_{\rm s}^D$ and $\rho_{\rm s}^V$ are the density part and vacuum fluctuation part of the scalar density, respectively. 
The $U^{\rm V,R}_{\rm 1-loop}$ is the renormalized effective potential induced by the 1-loop vacuum fluctuation effects and is given by  \cite{rf:Chin}
\begin{eqnarray}
U^{\rm V,R}_{\rm 1-loop}(<\phi >)=&-&{\lambda\over{8\pi^2}}[ M^{*4}\ln{\left( M^*/M \right)}+M^3(M-M^*)
\nonumber
\\-{7\over{2}}M^2(M-M^*)^2&+&{13\over{3}}M(M-M^*)^3-{25\over{12}}(M-M^*)^4 ]. 
\label{eq:4}
\end{eqnarray}

The energy density of the nuclear matter is also given by  
\begin{eqnarray}
\varepsilon (k_F,<\phi >,<V^0>)&=&\varepsilon_{\rm N}(k_F,M^*)+\varepsilon_{\rm v}(<V^0>)
\nonumber\\
&+&{1\over{2}}\mu_{\rm s}^2<\phi >^2+U^{\rm V,R}_{\rm 1-loop}(<\phi >), 
\label{eq:2} 
\end{eqnarray}
where 
\begin{equation}
\varepsilon_{\rm N}(k_F,<\phi >)={\lambda\over{12\pi^2}}\left[ 3k_F^3E_F^*+{3\over{2}}M^{*2}k_FE_F^*-{3\over{2}}\ln{\left( {k_F+E_F^*\over{M^*}}\right)} \right]
\label{eq:3} 
\end{equation}
and 
\begin{equation}
\varepsilon_{\rm v} (k_F,V_0)=g_{\rm v}<V^0>\rho -{\mu_{\rm v}^2\over{2}}<V^{0}>^2={\mu_{\rm v}^2\over{2}}<V^0>^2.  
\label{eq:5}
\end{equation}
It is easy to show that
\begin{equation}
\rho_{\rm s}^{\rm D}=-{1\over{g_{\rm s}}}{\partial \varepsilon_{\rm N}\over{\partial <\phi >}}
\label{eq:ad31}
\end{equation}

Using the thermodynamical identity, 
we obtain 
\begin{equation}
{\varepsilon + P\over{\rho}}=\mu=E_F^*+C_{\rm v}^2\rho, 
\label{eq:7}
\end{equation}
where $P$ and $\mu$ are the pressure and the baryonic chemical potential of the nuclear matter, respectively. 
At the normal density $\rho_0$, the pressure $P$ vanishes. 
Then, Eq. (\ref{eq:7}) yields \cite{rf:Boguta}
\begin{equation}
C_{\rm v}^2=(-a_1+M-\sqrt{k_{F0}^2+M_0^{*2}})/\rho_0,  
\label{eq:9}
\end{equation}
where the quantity with zero subscript shows the one at the normal density and  $a_1$ is the value of the binding energy. 
Equation (\ref{eq:9}) and the condition $C_{\rm v}^2 >0$ gives a condition 
$M^*_0/M<0.944$. 
The incompressibility $K$ of the nuclear matter is given by 
\begin{equation}
K=9\rho_0^2\left. {\partial ^2(\epsilon /\rho )\over{\partial \rho^2}}\right|_{\rho =\rho_0} 
=9\left. {\partial P\over{\partial \rho}}\right|_{\rho =\rho_0}=9\rho_0\left. {\partial \mu\over{\partial \rho}}\right|_{\rho =\rho_0}
\label{eq:10}
\end{equation}
Putting $\mu=E_F^*+C_{\rm v}^2\rho$ into (\ref{eq:10}), we obtain 
\begin{equation}
K=9\rho_0\left. \left( {k_F^3\over{3\rho E_F^*}}+{g_{\rm v}^2\over{\mu_{\rm v}^2}}+{M^*\over{E_F^*}}{dM^*\over{d\rho}}\right) \right|_{\rho =\rho_0}
\label{eq:11}
\end{equation}

Next we calculate the self-energy $\Pi_{\rm s}$ of the $\sigma$-meson by using the random phase approximation (RPA). \cite{rf:Furn}
Using the same renormalization conditions as (\ref{eq:ad3}) for the effective potential and the usual renormalization conditions for the $\sigma$-meson wave function, we obtain 
\begin{equation}
\Pi_{\rm s}(q;<\phi >,k_F)=\Pi_{\rm s}^V(q^2;<\phi >)+\Pi_{\rm s}^D(q;<\phi >,k_F)
\label{eq:21}.  
\end{equation}
The particle-antiparticle excitation part $\Pi_{\rm s}^V$ does not depend explicitly on $k_F$ and is given by 
\begin{equation}
 \Pi _{\rm s} ^V (q^2;<\phi >) = \frac{{3g_{\rm s} ^2 }}{{4\pi ^2 }}\int\limits_0^1 {{\rm d}x} \left[ {3M^{*2}  + M^2  - 4MM^*  - q^2 x(1 - x) - A^{*2} \ln \frac{{A^{*2} }}{{M^2 }}} \right]{\rm  ,} 
\label{eq:22}   
\end{equation}
where $A^{*2}  = M^{*2}  - q^2 x(1 - x)$. 
The $\Pi_{\rm s}^D$ which includes the particle-hole excitations and the Pauli-blocking effects depends explicitly on $k_F$ and is given by 
\begin{eqnarray}
\Pi _{\rm s} ^D(q;<\phi>,k_F) = &-&ig_{\rm s}^2\int \frac{{{\rm d}^4 k}}{{(2\pi )^4 }}
{\rm Tr}
[ G_D(k)G_F(k+q)
\nonumber\\
&+&G_F(k)G_D(k+q)+G_D(k)G_D(k+q)]
\label{eq:23}
\end{eqnarray}
where $G_D$ and $G_F$ are the density part and the Feyman part of the nucleon propagator in RHA. 

Now we define two kinds of effective $\sigma$-meson mass. 
First, we define an "off-shell" effective mass $\mu_{\rm s}^*$ by 
\begin{equation}
\mu_{\rm s}^{*2}\equiv \mu_{\rm s}^2 +\lim_{q_0\to 0}\lim_{\vert {\bf q}\vert \to 0}\Pi_{\rm s}(q). 
\label{eq:24} 
\end{equation}
The $\mu_{\rm s}^*$ can be regarded as a range of the nuclear force which is mediated by the $\sigma$-meson in the nuclear matter. 
The $\mu_{\rm s}^*$ can be related to the effective potential of the nuclear matter. 
In fact, it is easy to show that 
\begin{equation}
\mu_{\rm s}^{*2}={\partial^2 \varepsilon \over   {\partial <\phi >^2}}. 
\label{eq:25}
\end{equation}

Differentiating the equation of motion (\ref{eq:ad1}) for the $\sigma$-meson field with respect to the baryon density and using Eqs. (\ref{eq:ad31}) and (\ref{eq:25}), we obtain 
\begin{equation}
{dM^*\over{d\rho}}=-{{g_{\rm s}^2M^*}\over{E_F^*{\partial ^2 \epsilon\over{\partial <\phi>^2}}}}=-{g_{\rm s}^2\over{\mu_{\rm s}^{*2}}}{M^*\over{E_F^*}}
\label{eq:12}
\end{equation}
Putting (\ref{eq:12}) into (\ref{eq:11}), we obtain 
\begin{equation}
K=9\rho_0\left. \left( {k_F^3\over{3\rho E_F^*}}+{g_{\rm v}^2\over{\mu_{\rm v}^2}}-{g_{\rm s}^2\over{\mu_{\rm s}^{*2}}}{M^{*2}\over{E_F^{*2}}}\right) \right| _{\rho =\rho_0}. 
\label{eq:13}
\end{equation}

The $\mu_{\rm s}^*$ is not an "on-shell" mass which is defined by the pole of the propagator 
\begin{equation}
\Delta (q) = \frac{1}{{q^2  - (\mu_{\rm s} ^2  + \Pi _{
\rm s} (q))}}{\rm  }{\rm .}
\label{eq:26}
\end{equation}
We define the "on shell" effective mass $m_{\rm s}^*$ by the equation 
\begin{equation}
m_{\rm s}^{*2}\equiv \mu_{\rm s} ^2  + \left. {\Pi _{\rm s}  (q)} \right|_{\scriptstyle {\bf q}^2  = 0 \hfill \atop 
  \scriptstyle q_0 ^2  = m_{\rm s} ^{*2}  \hfill}. 
\label{eq:27} 
\end{equation}
In particular, at $\rho =0$, we define 
\begin{equation}
m_{\rm s}^{2}
\equiv  \mu_{\rm s} ^2  + \left. {\Pi _{\rm s}  (q)} \right|_{\scriptstyle {\bf q}^2  = 0 \hfill \atop 
  \scriptstyle q_0 ^2  = m_{\rm s} ^2  \hfill}
{\rm  }{\rm ,}
\label{eq:28}
\end{equation}
where $m_{\rm s}$ is the physical mass of $\sigma$-meson. 

In the ordinary RHA $K=473$MeV, which is much larger than the empirical value 150$\sim$350MeV. \cite{rf:Pearson,rf:Shlomo}  
Therefore, we add an additional potential of the $\sigma$-meson self-interaction
\begin{equation}
 U^{\rm H}(<\phi > ) = \sum_{n = 5}^\infty D_n (g_{\rm s} <\phi > )^n=\sum_{n=5}^\infty D_n(M-M^*)^n 
\label{eq:14}
\end{equation}
to the Lagrangian (\ref{eq:1}). 
We regard $U^{\rm H}(<\phi >)$ as the effective potential induced by the higher-order quantum corrections beyond 1-loop approximation. 
We remark that, in Eq. (\ref{eq:14}),  the terms of $<\phi >^0\sim <\phi >^4$ have been canceled by the counter term to the higher-order quantum corrections just as Eq. (\ref{eq:ad3}). 
By this modification, 
Eqs (\ref{eq:2}) and (\ref{eq:21}) are modified as 
\begin{eqnarray}
\varepsilon (k_F,<\phi >,<V^0>)&=&\epsilon_N(k_F,M^*)+\epsilon_{\rm v}(<V^0>)
\nonumber\\
+{1\over{2}}\mu_{\rm s}^2<\phi >^2&+&U^{\rm V,R}_{\rm 1-loop}(<\phi >)+U^{\rm H}(<\phi >) 
\label{eq:31} 
\end{eqnarray}
and 
\begin{equation}
\Pi_{\rm s}(q;<\phi >,k_F)=\Pi_{\rm s}^V(q^2;<\phi >)+\Pi_{\rm s}^D(q;<\phi >,k_F)+g_{\rm s} ^2 {d^2\over{d {M^{*}}^2}}U^H(<\phi >){\rm  .} 
\label{eq:32} 
\end{equation}
We remark that Eqs. (\ref{eq:25}) and (\ref{eq:13}) are still valid, after this expansion was carried out. 

\section{Numerical calculation}
Equation (\ref{eq:13}) gives the relation among the effective nucleon mass, the incompressibility and the effective $\sigma$-meson mass $\mu_{\rm s}^*$ at the normal density. 
If the values of $M^*_0$ and $K$ is given, $C_{\rm v}=g_{\rm v}/\mu_{\rm v}$ is determined by Eq. (\ref{eq:9}) and we can calculate the ratio $\mu_{\rm s}^*/g_{\rm s}$ at the normal density. 
In Fig. 1, we display the ratio $\mu_{\rm s}^*/g_{\rm s}$ as a function of $M^*_0$ with several values of $K$. 
In the numerical calculations, we set $a_1=$15.75MeV, $\rho_0=0.15$fm$^{-3}$ and $M=$939MeV. 
The ratio $\mu_{\rm s}^*/g_{\rm s}$ increases as $M^*_0$ increases, while the ratio depends on $K$ only slightly. 

Since $U^H$ does not appear explicitly in (\ref{eq:13}), the result in Fig. 1 is established regardless of the details of the potential form.  
However, to calculate $\mu_{\rm s}^*$ itself, we must determine $g_{\rm s}$. 
For this purpose, we assume that \cite{rf:Sakamoto} 
\begin{equation}
 U^{\rm H}(<\phi > ) = D_5(M-M^*)^5+D_6(M-M^*)^6. 
\label{eq:33}
\end{equation}
In this approximation, 
we have four parameters for RHA calculation, namely, $C_{\rm s}$, $C_{\rm v}$
, $D_5$ and $D_6$. 
As is seen in Eq. (\ref{eq:9}), $C_{\rm v}$ is determined if the value of $M_0^*$ is given. 
We have two conditions for the saturations at $\rho =\rho_0$. 
\begin{equation}
\varepsilon(\rho_0) =(M-a_1)\rho_0~~~~~~{\rm and }~~~~~~P(\rho_0)=0
\label{eq:33a}
\end{equation}
Therefore, if the value of $K$ is given, the remaining three parameters $C_{\rm s}$, $D_5$ and $D_6$ are determined. 

Using the same potential as (\ref{eq:33}), we obtain 
\begin{eqnarray}
\Pi_{\rm s}(q;<\phi >,k_F)&=&\Pi_{\rm s}^V(q^2;<\phi >)+\Pi_{\rm s}^D(q;<\phi >,k_F)
\nonumber\\
&+&20g_{\rm s}^2D_5(M-M^*)^3+30D_6g_{\rm s}^2(M-M^*)^4{\rm  ,} 
\label{eq:ad51} 
\end{eqnarray}
From Eqs. (\ref{eq:28}) and (\ref{eq:ad51}) , at zero density, we obtain 
\begin{equation}
1 + C_{\rm s}^2\left. {{\Pi _{\rm s} ^V (q)}\over{g_{\rm s}^2}} \right|_{\scriptstyle {\bf q}^2  = 0 \hfill \atop 
  \scriptstyle q_0 ^2  = m_{\rm s} ^2  \hfill}= \frac{{m_{\rm s} ^2 }}{{\mu_{\rm s} ^2 }}{\rm  }{\rm .}
\label{eq:34} 
\end{equation}
Since $\Pi_{\rm s}^V/g_{\rm s}^2$ does not depend on $g_{\rm s}$, we can calculate $\mu_{\rm s}$ by putting the value of $C_{\rm s}$ and $m_{\rm s}=$550MeV into Eq. (\ref{eq:34}). 
After determining $\mu_s$, we can also determine the value of $g_s$. 
In Figs. 2 and 3, we display $\mu_{\rm s}$ and $g_{\rm s}$ as a function of $M^*_0$. 
As $M^*_0$ increases, $g_{\rm s}$ is suppressed more strongly than $\mu_{\rm s}$. 
Since, as is seen in (\ref{eq:9}), $C_v^2$ approches zero as $M^*_0\to 0.944M$, $g_s$ becomes small more quickly than $\mu_{\rm s}$ to keep the saturation conditions. 
Both of $\mu_{\rm s}$ and $g_{\rm s}$ depend on $K$ only slightly. 

After determining $\mu_{\rm s}$ and $g_{\rm s}$, we can calculate $\mu_{\rm s}^*$ and $m_{\rm s}^*$. 
In Figs. 4 and 5, we display $\mu_{\rm s}^*/\mu_{\rm s}$ and $m_{\rm s}^*/m_{\rm s}$ at the normal density as a function of $M^*_0$. 
As $M^*_0$ increase, the ratios $\mu_{\rm s}^*/\mu_{\rm s}$ and $m_{\rm s}^*/m_
{\rm s}$ increase. 
The ratio $\mu_{\rm s}^*/\mu_{\rm s}$ is smaller than 1 for $M^*_0<0.560(0.542, 0.526)M$, when $K=$200(300,400)MeV.  
Similarly, the ratio $m_{\rm s}^*/m_{\rm s}$ is smaller than 1 for $M^*_0<0.815(0.802,0.793)M$, when $K=$200(300,400)MeV. 
Both of two ratios depend on $K$ only slightly. 

In Figs. 6 and 7, we display $\mu_{\rm s}^*/\mu_{\rm s}$ and $m_{\rm s}^*/m_{\rm s}$ as a function of baryon density with several values of $M^*_0$. 
Since two ratios depend on $K$ only slightly, we have fixed the value of $K$ at 300MeV. 
The density dependence of the two ratios also changes, when $M^*_0$ changes. 
The ratio $\mu_{\rm s}^*/\mu_{\rm s}~^>_\sim 1$ in the region of $\rho^<_\sim 1.3\rho_0$, while  the ratios $m_{\rm s}^*/m_{\rm s}< 1$ except for the case with $M_0^*=0.85M$.  

\section{Summary}
In summary, we have studied the relation among the effective nucleon mass $M^*$, the incompressibility $K$ and the effective $\sigma$-meson mass $m_{\rm s}^*$ in nuclear matter by using the relativistic nuclear model. 
We found that there is a strong correlation between $M^*$ and $m_{\rm s}^*$, while there is only a weak correlation between $K$ and $m_{\rm s}^*$. 
At the normal density, $m_{\rm s}^*$ is smaller than the one at zero density for $M^*_0\leq 0.8M$, while $\mu_{\rm s}^*$ hardly decreases. 
We remark that it is interesting that the off-shell mass $\mu_{\rm s}^*$ at the normal density is related directly to $M^*_0$ and $K$ which can be determined phenomenologically. 

\bigskip

\centerline{\bf ACKNOWLEDGEMENTS}

The authors thanks Prof. T. Kohmura and T. Maruyama for useful discussions. 


\bigskip
\bigskip
\begin{center} {\large{\bf{Figures}}}
\end{center}
%
\begin{figure}[ht]
\center\epsfxsize=8cm\leavevmode\epsffile{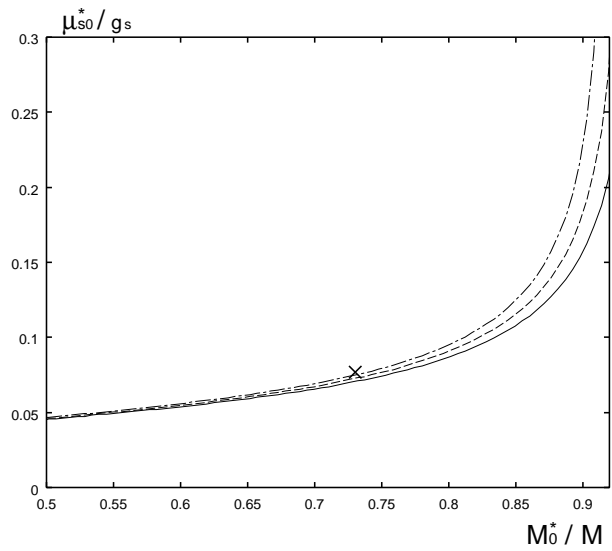}
\caption{The ratio $\mu_{s0}^*/g_{\rm s}$ as a function of $M^*_0$. 
The solid, the dashed and dashed-dotted curves show results for $K=$200MeV, 300MeV and 400MeV, respectively. 
The cross show the result for the original RHA.}
\label{fig1}
\end{figure}
%
\begin{figure}[ht]
\center\epsfxsize=8cm\leavevmode\epsffile{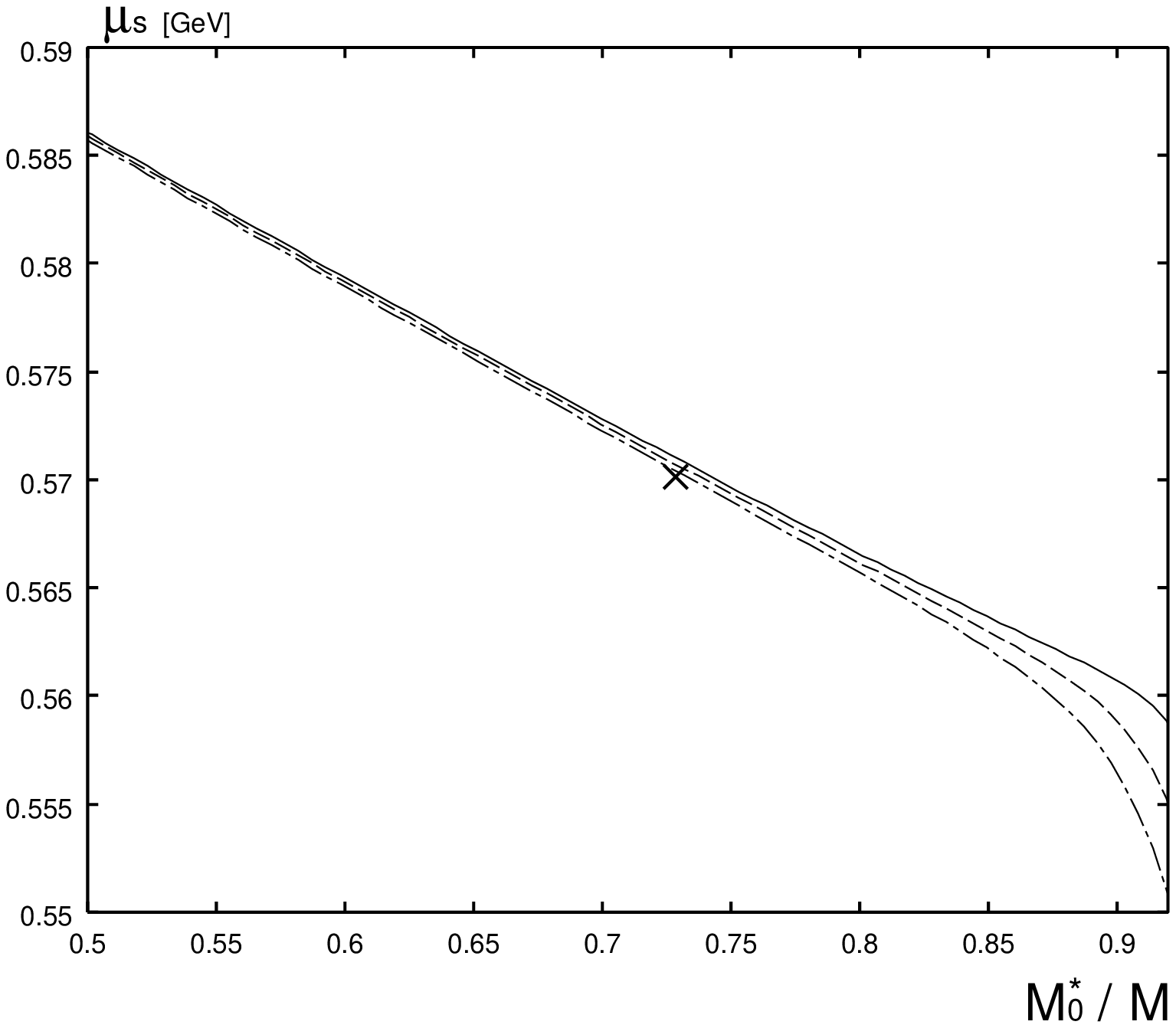}
\caption{The $\mu_{\rm s}$ as a function of $M^*_0$. 
The various curves and the cross have the same notation as in Fig. 1.}
\label{mst}
\end{figure}
%
\begin{figure}[ht]
\center\epsfxsize=8cm\leavevmode\epsffile{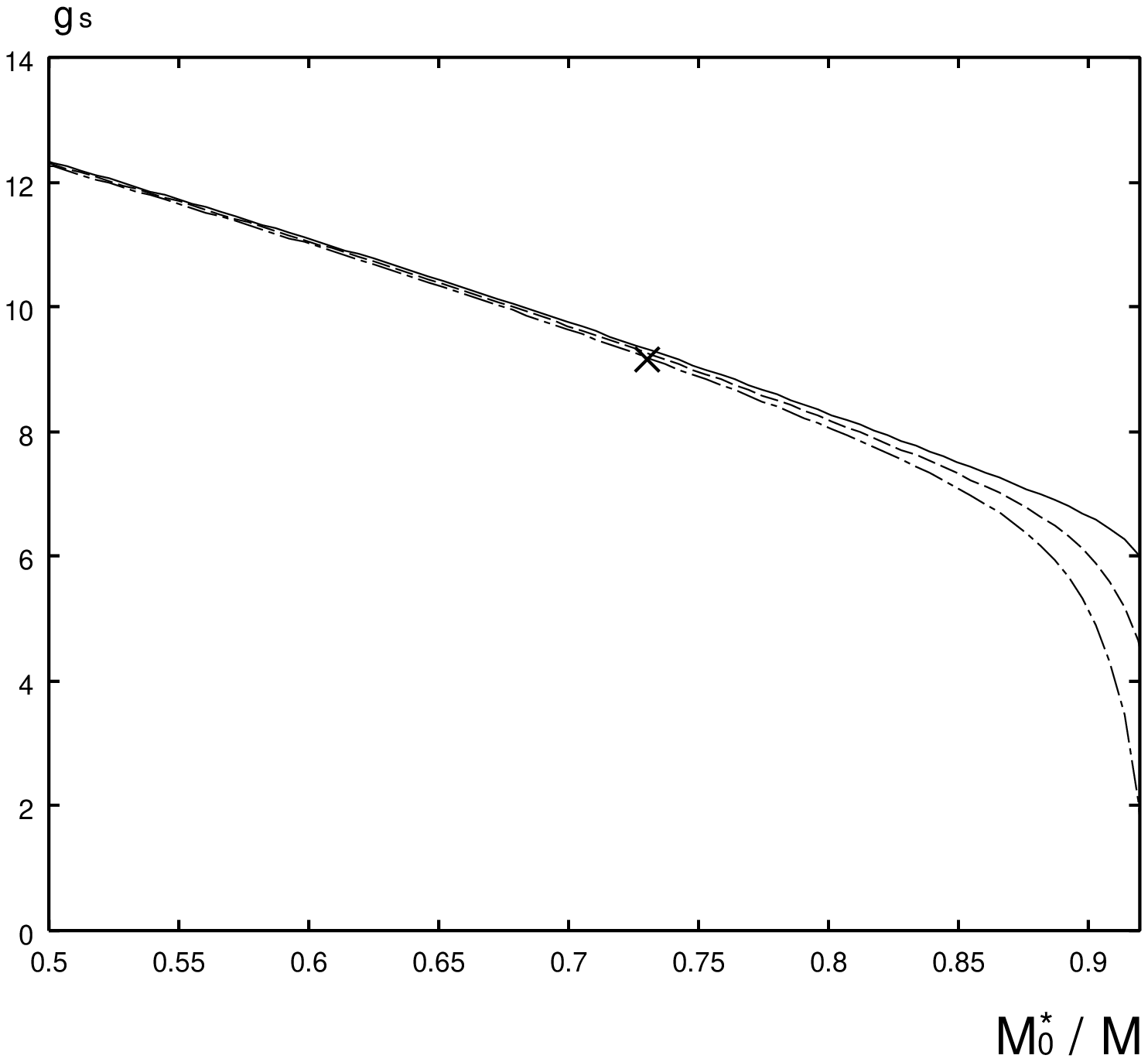}
\caption{The $g_{\rm s}$ as a function of $M^*_0$. 
The various curves and the cross have the same notation as in Fig. 1}
\label{gs}
\end{figure}
%
\begin{figure}[ht]
\center\epsfxsize=8cm\leavevmode\epsffile{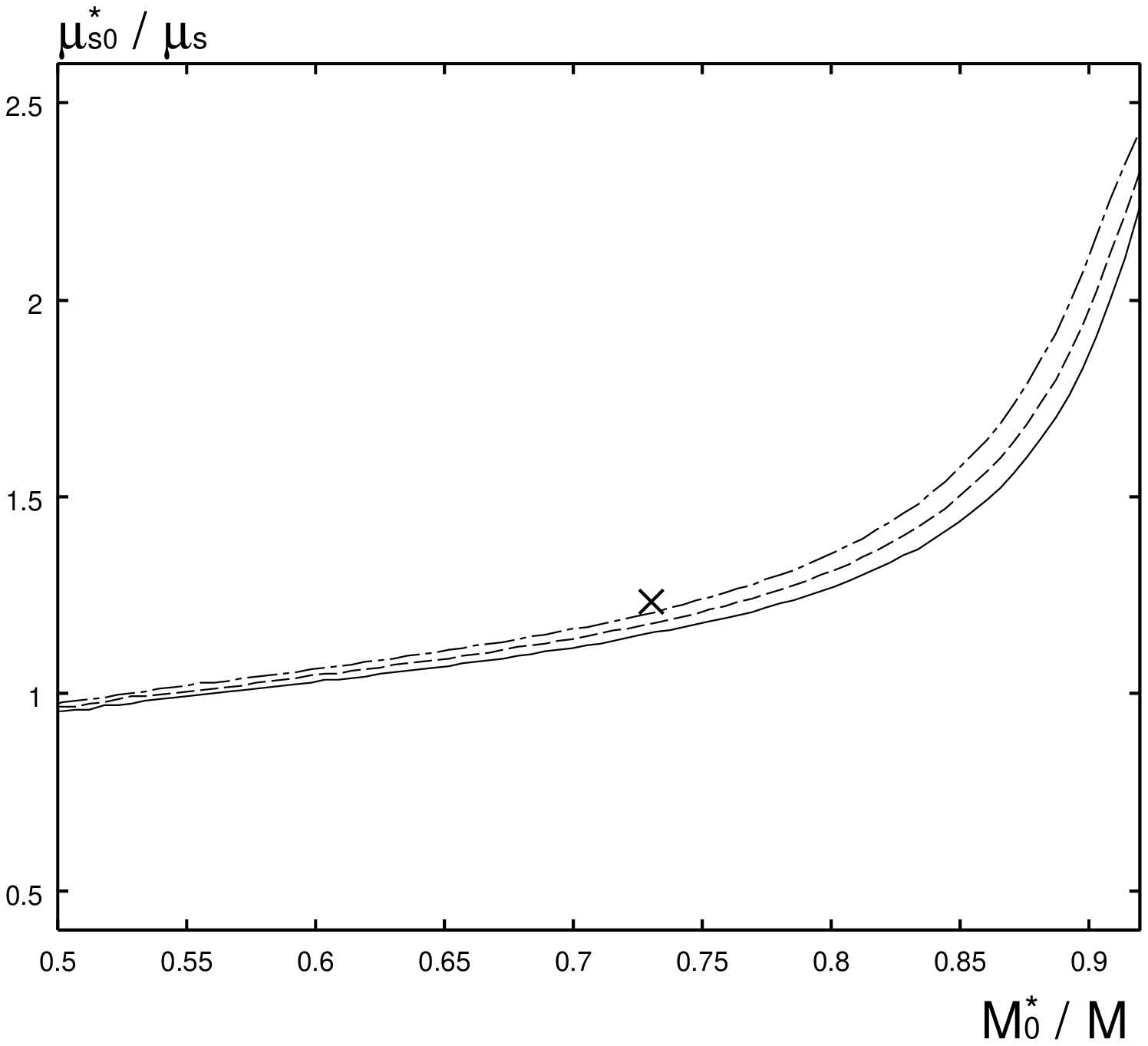}
\caption{The ratio $\mu_{s0}^*/\mu_{\rm s}$ as a function of $M^*_0$. 
The various curves and the cross have the same notation as in Fig. 1.} 
\label{fig4}
\end{figure}
%
\begin{figure}[ht]
\center\epsfxsize=8cm\leavevmode\epsffile{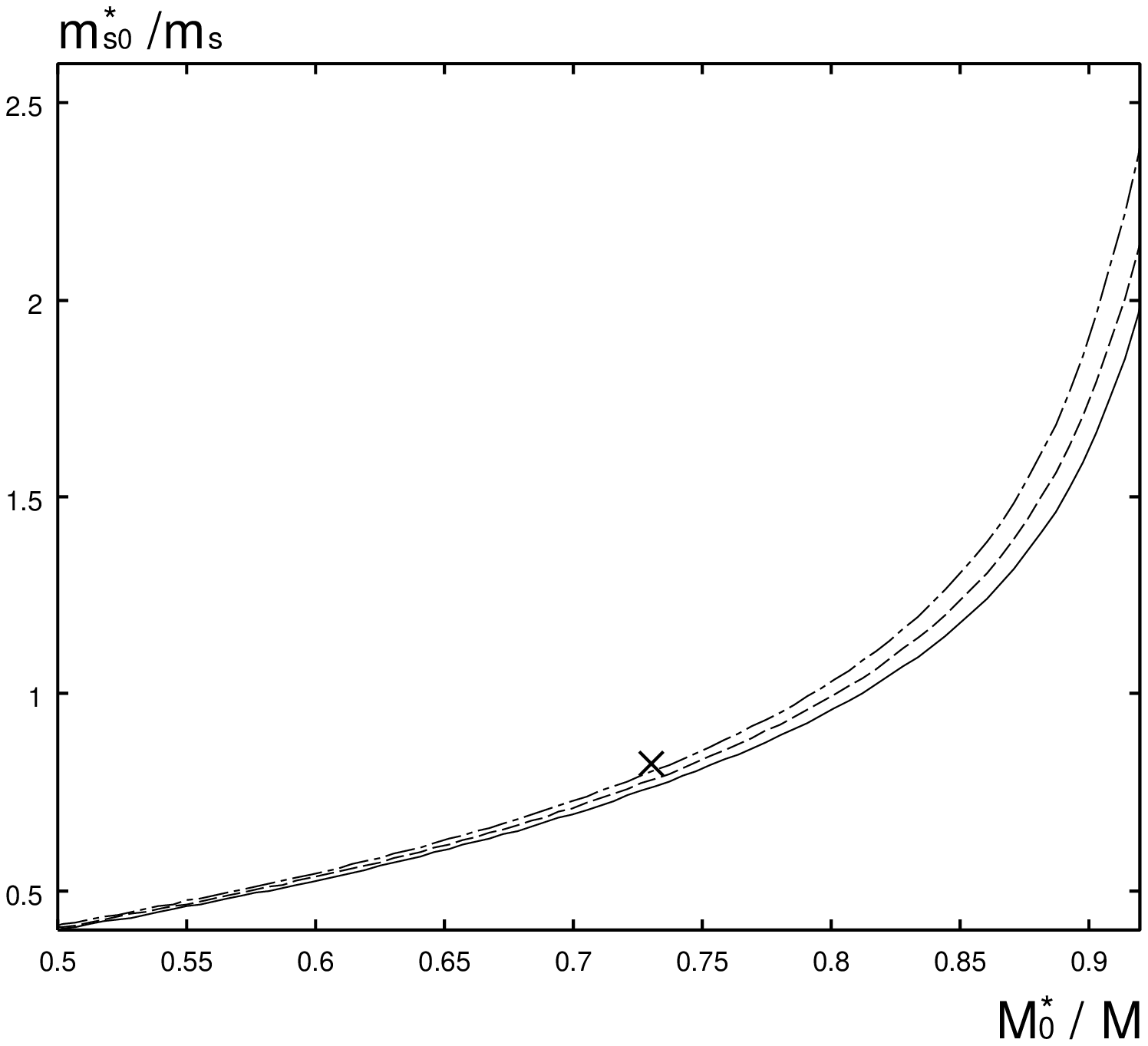}
\caption{The ratio $m_{s0}^*/m_{\rm s}$ as a function of $M^*_0$. 
The various curves and the cross have the same notation as in Fig. 1.} 
\label{fig5}
\end{figure}
%
\begin{figure}[ht]
\center\epsfxsize=8cm\leavevmode\epsffile{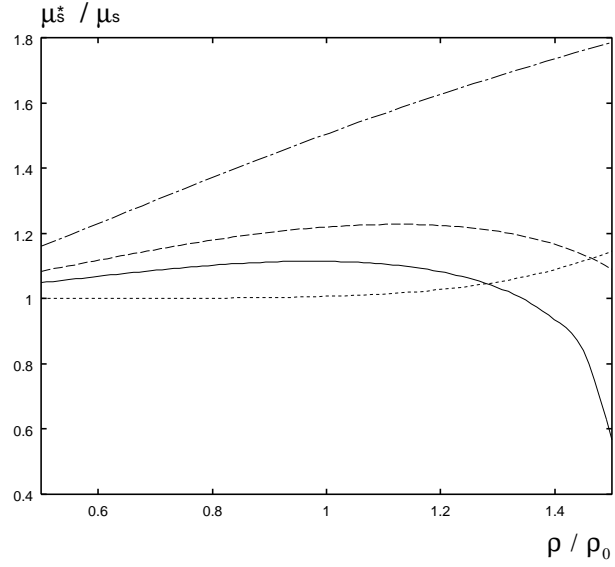}
\caption{The ratio $\mu_{\rm s}^*/\mu_{\rm s}$ as a function of the baryon density. 
The solid, the dashed and dashed-dotted curves show results for $M_0^*=0.65M$, $0.75M$ and $0.85M$, respectively. 
The dotted curve shows the result for the original RHA.} 
\label{fig6}
\end{figure}
%
\begin{figure}[ht]
\center\epsfxsize=8cm\leavevmode\epsffile{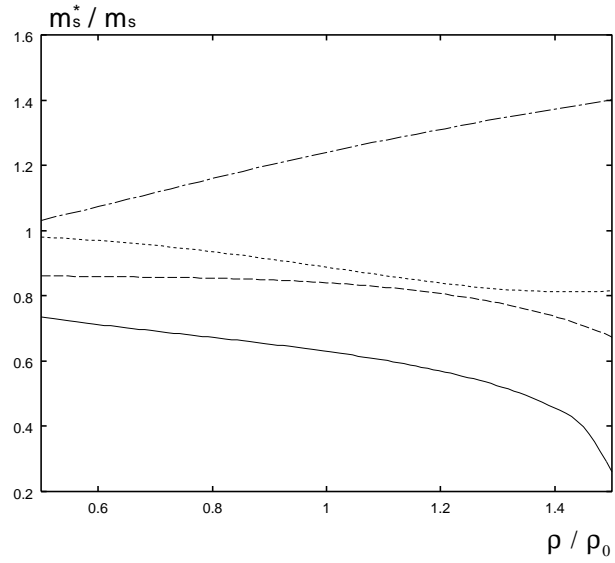}
\caption{The ratio $m_{\rm s}^*/m_{\rm s}$ as a function of 
the baryon density. 
The various curves have the same notation as in Fig. 6.} 
 \label{fig7}
\end{figure}
%

\end{document}